\documentclass[11pt]{article}  
\usepackage{cancel} 
\usepackage{simplewick} 
\usepackage{fullpage} 	
\usepackage{amssymb}     
\usepackage{amsmath}
\usepackage{graphicx,subcaption} 
\usepackage{multirow} 
\usepackage{dsfont} 
\usepackage{cite}       
\usepackage{axodraw4j}
\usepackage{titling}

\newcommand\e{\epsilon}
\renewcommand\k{\kappa}
\newcommand\g{\gamma}
\renewcommand\d{\partial}

\renewcommand\>{\rangle}
\renewcommand\O{{\mathcal O}}

\begin{document}

\pretitle{\begin{flushright}\vspace{-85pt}INT-PUB-12-034\end{flushright}}
\title{}
\posttitle{\begin{center} \LARGE (Non)-Renormalization of the Chiral Vortical Effect Coefficient \end{center}} 
\author{Siavash Golkar and Dam T. Son\\
	 \normalsize{\textit{Institute for Nuclear Theory, University of Washington, Seattle, WA 98195-1550, USA}}}


\maketitle

\begin{abstract}
We show using diagramtic arguments that in some (but not all) cases, the temperature dependent part
of the chiral vortical effect coefficient is independent of the
coupling constant. An interpretation of this result in terms of
quantization in the effective 3 dimensional Chern-Simons theory is
also given. In the language of 3D, dimensionally reduced theory, the
value of the chiral vortical coefficient is related to the formula
$\sum_{n=1}^\infty n=-1/12$. We also show that in the presence of dynamical gauge fields, the CVE coefficient is not protected from renormalization, even in the large $N$ limit.
\end{abstract}

\section{Introduction}

Recently, the manifestation of chiral anomaly in hydrodynamics has
been under considerable attention.  Since the original
observations in gauge/gravity
duality~\cite{Bhattacharyya:2007vs,Erdmenger:2008rm,Banerjee:2008th},
it has been shown that the hydrodynamic equations of
systems with anomalies necessarily contain additional terms if both the
equations of anomalies and the second law of thermodynamics are to be
preserved~\cite{Son:2009}.  These new terms lead to two new effects:
the chiral magnetic effect (CME) and the chiral vortical effect (CVE),
i.e. the appearance of a current in the presence of a magnetic field
or a vorticity of a fluid flow.  These effects are dissipationless and
hence are proportional to equilibrium thermodynamic quantities rather
than kinetic coefficients~\cite{Banerjee:2012,Jensen:2012jy}.

It has been found that anomaly matching within hydrodynamics fixes the
coefficents of the CME and CVE up to a single numerical
coefficient~\cite{Neiman:2010,Son:2009}.  This coefficient determines the
magnitude of the CVE at zero chemical potential.  Consider a fluid
flow with a velocity profile $u^\mu(x)$.  The coefficient under
consideration is $C_1$ in the following relationship
\begin{equation}
  j^{5\mu} = C_1 T^2 \frac12\epsilon^{\mu\nu\lambda\rho} u_\nu\d_\lambda u_\rho.
\end{equation}
Landsteiner et al. \cite{Landsteiner:pert,Landsteiner:holo} show that in
theories with gravity dual, the value of this coefficient in
holography (i.e. at strong coupling) coincides with its value at zero
coupling, suggesting coupling-constant independence (For a different approach based on kinetic theory see \cite{Gao:2012ix}).  In a theory with
a single fermion of right-handed chirality, $C_1=1/12$.  Landsteiner
et al. suggested that this coefficient is related to a gravitational
anomaly (more precisely, the gravitational contribution to the
divergence of an axial current).  Although the connection between the
zero-chemical-potential CVE and the gravitational anomaly is direct in
holography, it has not been proven outside of holography. In fact,
naive power counting indicates that the effects of gravitational
anomaly can appear only in higher-order (namely, third-order)
hydrodynamics.

In this paper, we demonstrate that in a system of fermions coupled 
through Yukawa interactions,  the CVE coefficient at zero chemical
potential is independent of the coupling-constant.  We
first provide a perturbative analysis of diagrams to show that the 
one-loop fermion graph is the only graph contributing to the relevant
Green function appearing in the Kubo's formula for zero-$\mu$ CVE 
coefficient.  We then show a connection between the CVE and the 3D
Chern-Simons term appearing in the dimensionally reduced effective
field theory of the thermal system.  Using this connection, we show
that we can understand the numerical value of $C$ from the summation
formula $\sum_{n=1}^\infty n=-1/12$. (The non-renormalization in this case can also be inferred from the results in \cite{Nair:2012}, where a different approach based on fluid dynamics in terms of group-valued variables was taken.)

However, it turns out that for the case of coupling to gauge fields, the CVE coefficient is not in fact protected against renormalization. We demonstrate that even in the large $N$ limit, there is a single class of diagrams that contribute to the CVE coefficienct and explicitly evaluate this at the two loop order.

The paper is organized as follows.  In Sec.~\ref{sec:Coleman_Hill} we
present a perturbative proof of the non-renormalization of the chiral
vortical coefficient $C_1$.  Our proof is based on similar proof, by
Coleman and Hill, of the non-normalization of the fermion-induced
Chern-Simons terms in (2+1) dimensions.  We elaborate more on the
connection to the connection to the 3D Chern-Simons theory in
Sec.~\ref{sec:CS}.  We conclude in Sec.~\ref{sec:concl}. 

\emph{Note:} As this work was nearing completion, we were informed of \cite{Kristan:2012} which deals with issues similar to the ones considered in this paper. The first version of this paper included an error which was kindly pointed out by H. Ren et al. who also have a forthcoming paper discussing its implications\cite{Ren:2012}.

\section{Calculation of the CVE coefficient}
\label{sec:Coleman_Hill}
In this section, we look at two cases. First we consider a Dirac fermion coupled to scalars through a chiral Yukawa term and show that the CVE is independent of this ineraction. In the second part of his section, we show that this is also the case for fermions coupled to a dynamical gauge field in leading order in the large $N$ expansion, where $N$ is the number of colors associated with the gauge field. 

\subsection{Coupling To Scalars}
To be concrete, we first concentrate on a simplest interacting
quantum field theory with a conserved axial current; namely, a linear sigma
model which contains a Dirac fermion interacting with a scalar field.
\begin{multline}
S=\int\!  dt \,d^3x \, \left[i\bar{\psi}\gamma^\mu D_\mu \psi 
  - D_\mu\phi^* D^\mu\phi - m^2 \phi^*\phi - \lambda(\phi^*\phi)^2
  - g\left( \bar\psi \frac{1+\gamma^5}2 \psi \phi^*
     + \bar\psi \frac{1-\gamma^5}2 \psi \phi \right)\right.\\
   \left.  - \frac12 h_{\mu\nu}T^{\mu\nu}+ \O(h^2)\right],
\label{eq:Fermion_L}
\end{multline}
with
\begin{align}
 D_\mu \psi =& (\d_\mu-i\gamma^5A_\mu)\psi,\notag\\
 D_\mu \phi =& (\d_\mu -2iA_\mu)\phi,\notag\\
 T^{\mu\nu}=&\frac{i}{4}\bar\psi\g^\mu(\stackrel{\rightarrow}{D^\nu}
 -\stackrel{\leftarrow}{D^\nu})\psi+\mu\leftrightarrow\nu
  + \textrm{scalar contributions}.
\end{align}
We have include in the action the coupling to external axial U(1)
gauge field $A_\mu$ and small metric perturbation $h_{\mu\nu}$. 

The symmetries associated with the background fields are anomalous, and the
anomalies are captured by the Ward identities:
\begin{align}
 \d_\mu J^\mu_5=& 
    -\frac1{48\pi^2} \e^{\mu\nu\rho\sigma} F_{\mu\nu}F_{\rho\sigma},\\
 \d_\mu T^{\mu\nu}=& F^{\nu\rho} \big(J_\rho - \frac1{12\pi^2} 
   {\e_\rho}^{\sigma\alpha\beta} A_{\sigma}F_{\alpha\beta}\big),
\label{eq:4d-ward}
\end{align}
where $J^\mu_5=\bar\psi\g^\mu\g^5\psi$, $F_{\mu\nu}$ is the field
strength associated with $A_\mu$. We note that
the axial current is conserved in the absence of
sources. 

The goal here is to calculate the chiral vortical coefficient at zero
chemical potential ($A_0=0$).  As explained in
Refs.~\cite{Landsteiner:holo}, this coefficient is not a proper kinetic
coefficient, but is basically an equilibrium quantity given by the
behavior of the retarded two-point Green's function between the
current $J_A^i$ and the momentum density $T^{0j}$ at zero frequency
and small momenta,
\begin{equation}
  G_R^{i,0j}(\omega, k)\Big|_{\omega=0} = i\epsilon_{ijn} k_n \sigma_A^\mathcal V
  + O(k^2)
  \label{eq:CVE_def}
\end{equation}
The fact that we are interested in the zero-frequency limit allows the
chiral vortical coefficient to be related to an Euclidean Green's function
\begin{equation}
i G_R^{i,0j}=\mathcal G^{i,0j}=\frac{\delta}{\delta A_i}\frac{\delta}{\delta g_{oj}}Z = \<J_A^i T^{0j}\> + \text{contact terms},
\label{eq:CVE-Greens}
\end{equation}
where $\mathcal G^{i,0j}$ is the Euclidean (Matsubara) Green's
function.  At zero frequency we can take our source fields $A_\mu$ and
$h_{\mu\nu}$ to be time independent from the start. With this
configuration, it is easy to see that the right hand side of the
anomalous Ward identities in equations \eqref{eq:4d-ward} vanish and
the symmetry is effectively restored.
 
We will demonstrate diagramatically, using an argument similar to that
of Coleman and Hill who showed the non-normalization of the
Chern-Simons term obtained by integrating out massive fermions in
3D\footnote{It is known that there are caveats to the Coleman-Hill
  theorem (see for example \cite{Khare:1994yv}). Namely when there are
  parity-violating interactions in the initial Lagrangian. However, as
  there are no such terms, they do not concern the treatment in this
  paper.},  that only diagrams that are present in the
zero-coupling limit give non-zero contributions to the CVE
coefficient\cite{Coleman-Hill}. The calculation at zero coupling is
given in \cite{Landsteiner:pert}, and our argument would show that the
calculation is exact at any value of the coupling constant.

\begin{figure}
  \centering
	\begin{subfigure}[b]{0.45\textwidth}
			\centering
			\includegraphics[width=\textwidth]{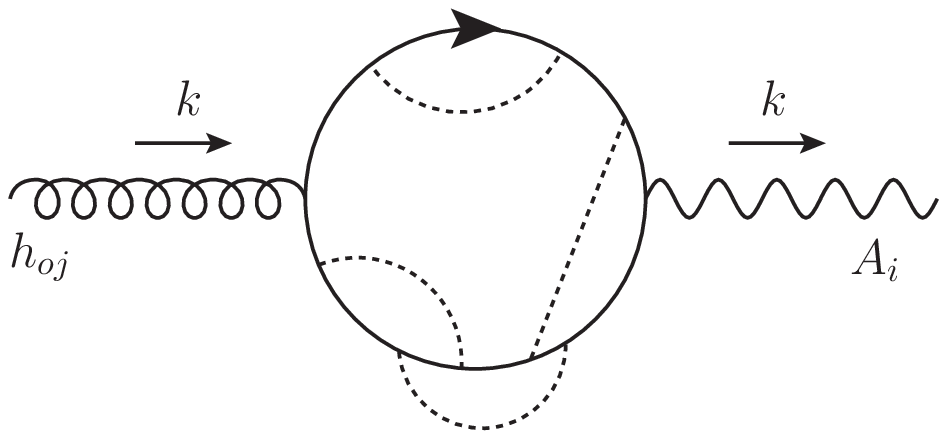}
			\caption{A generic diagram.}
			\label{fig:generic_diagram}
	\end{subfigure}
	\begin{subfigure}[b]{0.45\textwidth}
			\centering
			\includegraphics[width=\textwidth]{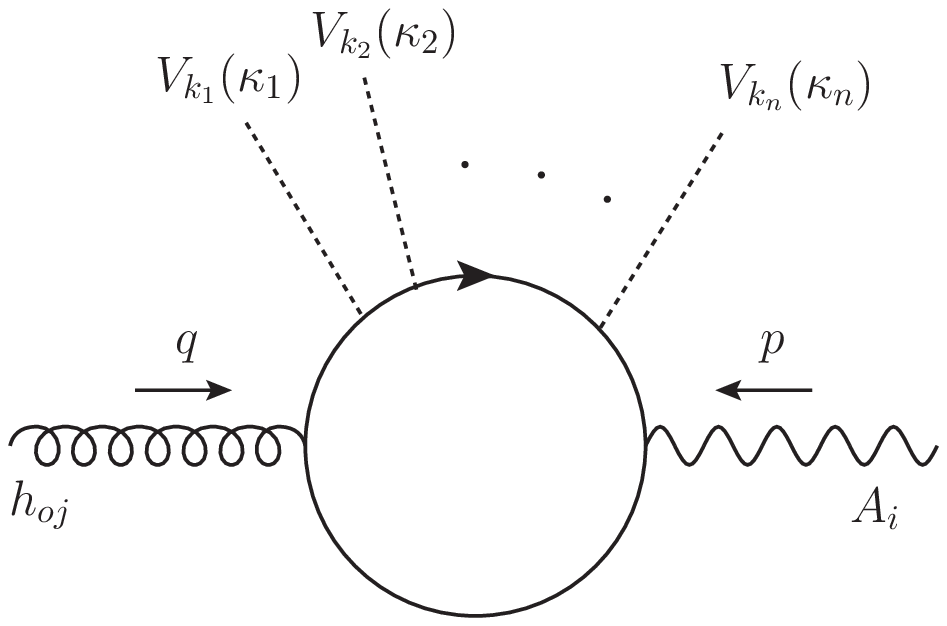}
			\caption{$n$-scalar effective vertex.}
			\label{fig:effective_vertex}
	\end{subfigure} 
  \caption{(a) A generic diagram at finite coupling. The internal lines are the sccalar field and the external lines are the graviton and the axial $U(1)$ field. (b) An $n$ scalar effective vertex. We can get all diagrams of type (a) by contracting the external scalar lines on type (b) diagrams.}
  \label{fig:finite_coupling}
\end{figure}

Let us now look at a generic diagram at non-zero coupling (see figure \ref{fig:generic_diagram}).  
All such diagrams can be obtained from the $n$-scalar effective vertices 
with exactly one insertion of $J^\mu_5$ and $T_{\mu\nu}$ by contracting
external scalar potentials and integrating over the scalar momenta.  Let
us look at the central block of the construction:

\begin{equation}
\Gamma^{(n)}_{ij}(p,q,\k_1,\cdots,\k_n).
\end{equation}
Graphically, this is a one-loop graph with $n$ external dynamical
scalars, 1 external graviton $b_i$ and 1 external source gauge boson
$a_i$ (figure \ref{fig:effective_vertex}).

The gauge invariance of the time independent source fields implies:
\begin{align}
p^i \Gamma^{(n)}_{ij}(p,q,\k_1,\cdots,\k_n)=0,\notag\\
q^j \Gamma^{(n)}_{ij}(p,q,\k_1,\cdots,\k_n)=0.
\label{eq:transverse_Gamma}
\end{align}
Differentiating these equations with respect to $p^r$ and $q^r$ and
then letting $p$ and $q$ be zero respectively we get:
\begin{align}
\label{eq:ward}
\begin{aligned}\Gamma^{(n)}_{ijk_1\cdots k_n}(0,q,\k_1,\cdots,\k_n)=0 \\
\Gamma^{(n)}_{ijk_1\cdots k_n}(p,0,\k_1,\cdots,\k_n)=0 \end{aligned}
\;\;\Rightarrow
\begin{aligned}\Gamma^{(n)}_{ijk_1\cdots k_n}(p,q,\k_1,\cdots,\k_n)=\O(p) \\
\Gamma^{(n)}_{ijk_1\cdots k_n}(p,k,\k_1,\cdots,\k_n)=\O(k), \end{aligned}
\end{align}
where we have used the fact that these functions are analytic in $k$. For $n=0$ the two momenta $p$ and $q$ are not independent and hence we derive:
\begin{align}
\Gamma^{(0)}_{ijk_1\cdots k_n}(p,-p)=\O(p).
\end{align}
However, for $n>0$ these two momenta are independent and \eqref{eq:ward} implies:
\begin{align}
\Gamma^{(n)}_{ijk_1\cdots k_n}(p,q,\k_1,\cdots,\k_n)=\O(pq).
\label{eq:ward_2nd}
\end{align}
This is all we need for our proof. The full contribution to the CVE coefficient is the sum of the zero coupling diagram calculated above and  diagrams with internal scalars. The latter can be constructed from our effective $n$ point vertices by contracting the external scalar lines.

Two cases arise. If the source fields $a_i$ and $b_i$ are attached to the same fermion loop, the total diagram is $\O(k^2)$ by equation \eqref{eq:ward_2nd}. Using the same argument, if they are attached to two different fermion loops, we again get that the total diagram is $\O(k^2)$ by applying equation \eqref{eq:ward} for each loop.  The same reasoning can be applied to the case with the scalar stress tensor insertion and the case with scalar internal loops, implying that all such diagrams are $\O(k^2)$. On the other hand the chiral vortical coefficient is the term
linear in the external momentum of the two-point Green's function.  Thus,
only the one-loop free diagram can contribute to the CVE. 
This completes the proof. 



It is important to note that the analyticity arguments given above require the mass of the scalars to be non-zero. Given the fact that scalars generically acquire mass at finite temperature, this requirement is equivalent to the statment that we are not sitting at a second order phase transition.

\subsection{Coupling To Gauge Fields}

We now turn to the case with fermions coupled to dynamical gauge fields. As claimed in the introduction, in this case, the CVE in fact does receive corrections. We show that for the non-abelian theory at leading order in $1/N$ there is a single class of diagrams that give non-zero contributions to the CVE coefficient. To demonstrate this, we point out where the arguments of the previous section are affected. The action is:
\begin{equation}
S=\int  dt \,d^3x \, \big(i\bar{\psi}\cancel D \psi -\frac1{4g^2} V_{\mu\nu}V^{\mu\nu}+ e A_\mu J^\mu_5 - \frac12 h_{\mu\nu}T^{\mu\nu}+\O(h^2) \big),
\label{eq:gauge_action}
\end{equation}
where $D$ and $V_{\mu\nu}$ are respectively the covariant (vector-like) coupling and curvature tensor associated with the dynamical gauge field $V_\mu$. We assume one flavor of fermions and we suppress the Lie group indices.

Following the steps of the scalar-coupling case discussed above, we come across two issues and we tackle them one by one, one of which ultimately leads to radiative corrections. The first difference  is that there are now potentially gapless excitations associated with the gluons. The infrared singularities associated with such massless modes threaten to ruin the analyticity arguments that were crucial in the proof. This is the same caveat that also came up in the original Coleman and Hill paper\cite{Coleman-Hill}. However, since we are working with non-abelian theories, the solution provided there does not apply. 

Fortunately, non-abelian theories at finite temperature do produce a mass gap non-perturbatively \cite{GPY}. To take advantage of this fact, we divide the arguments of the previous section into two steps. First, we perturbatively integrate out the fermionic fields to give effective vertices with gauge fields, the axial source and the graviton (respectively $V_\mu$, $A_i$ and $h_{0i}$) as external lines. According to the previous section, the diagrams with more than two external legs are $\O(k^2)$.

Next, we evaluate the expectation value of these effective vertices non-perturbatively. This method allows us to use both our perturbative diagramatic arguments of the previous section while steering clear of the IR singularities that plague perturbative analysis of gauge fields.

The second difference is that the anomaly (equation \eqref{eq:4d-ward}) no longer vanishes in the absence of external sources and is now given by:
\begin{equation}
\d_\mu J^\mu_5=-\frac{g^2C(r)}{16\pi^2\sqrt{-g}}\e^{\mu\nu\rho\sigma} V_{\mu\nu}V_{\rho\sigma}+\cdots,
\label{eq:vector_ward}
\end{equation}
where we have ommited the terms higher order in the external sources. This is a big problem as the arguments of the previous section all depended on the conservation of the axial current.

\begin{figure}[h]
  \centering
	\begin{subfigure}{0.48\textwidth}
			\centering
			\includegraphics[width=\textwidth]{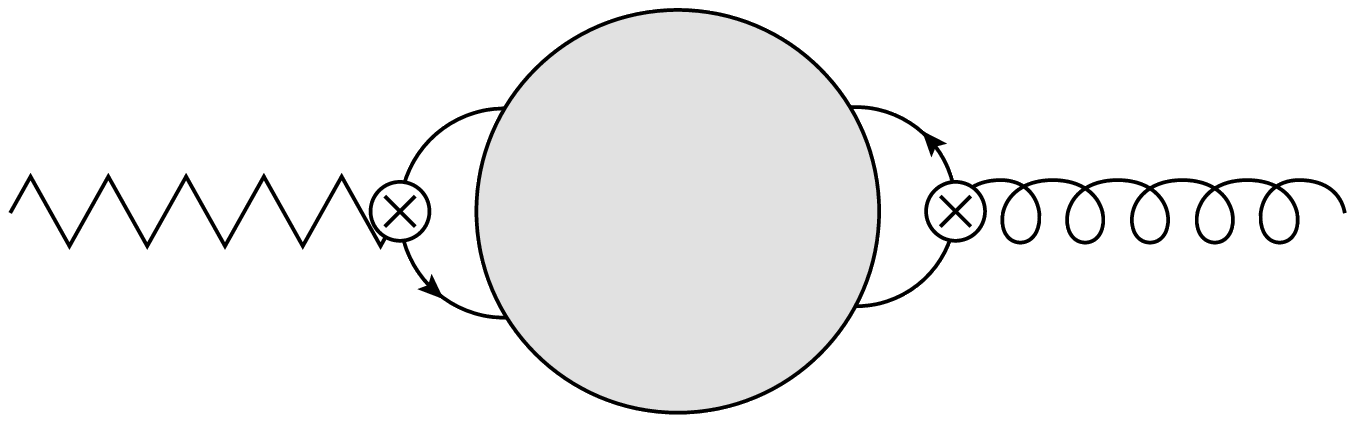}
			\caption{Fermionic stress tensor diagrams.}
			\label{fig:ferm_class}
	\end{subfigure}
	\;
	\begin{subfigure}{0.48\textwidth}
			\centering
			\includegraphics[width=\textwidth]{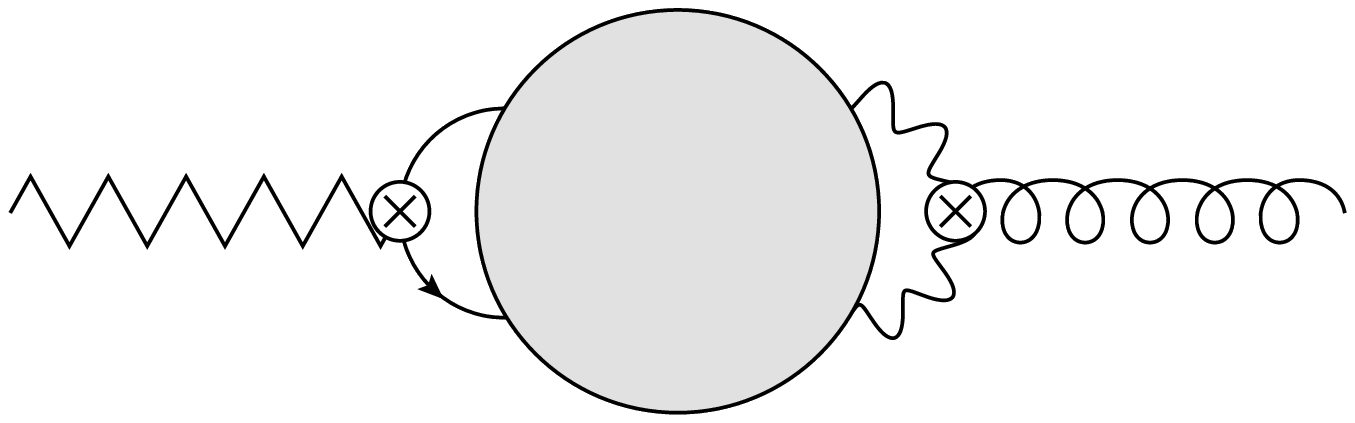}
			\caption{Gauge stress tensor diagrams.}
			\label{fig:gauge_class}
	\end{subfigure}
  \caption{The two diagram classes.}
  \label{fig:class}
\end{figure}

At this point, we split the diagrams into two groups depending on the form of the Energy-Momentum tensor (see figure \ref{fig:class}). The first group has a fermionic Energy Momentum insertion (figure \ref{fig:ferm_class}) and the second has the gauge part (figure \ref{fig:gauge_class}). For the first group, at leading order in $1/N$, we have a single fermion loop to which both the axial current $J^i_5$ and the  energy-momentum tensor $T^{0i}$ are attached. The anomalous contribution of such one loop diagrams is exactly captured in equation \eqref{eq:vector_ward}. Expanding the right hand side in metric perturbations, we see that to linear order, the anomaly only depends on the trace of the perturbation and thus does not contribute to correlators involving a single insertion of $T^{0i}$.

\begin{figure}[h]
  \centering
	\begin{subfigure}[b]{0.32\textwidth}
			\centering
			\includegraphics[width=\textwidth]{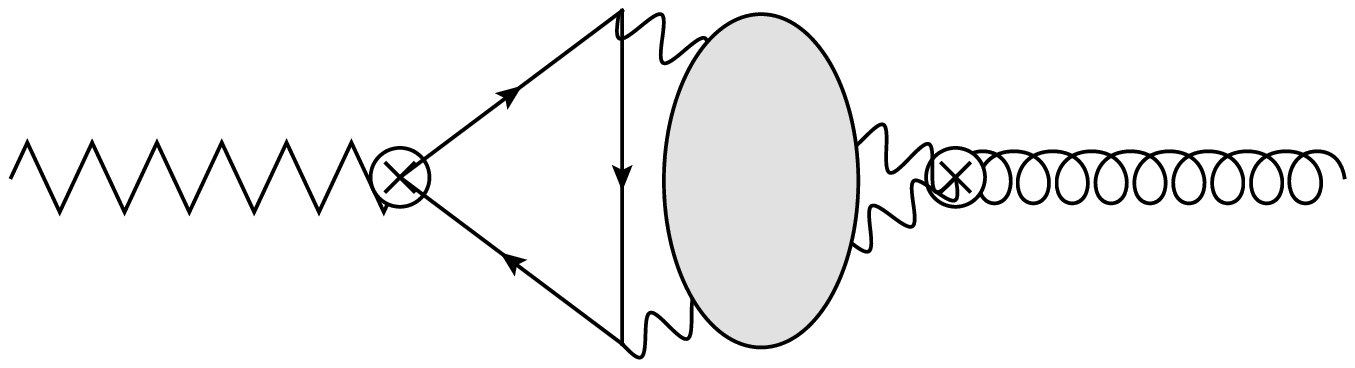}
			\caption{Anomalous diagrams.}
			\label{fig:AnomClass}
	\end{subfigure}
	\begin{subfigure}[b]{0.32\textwidth}
			\centering
			\includegraphics[width=\textwidth]{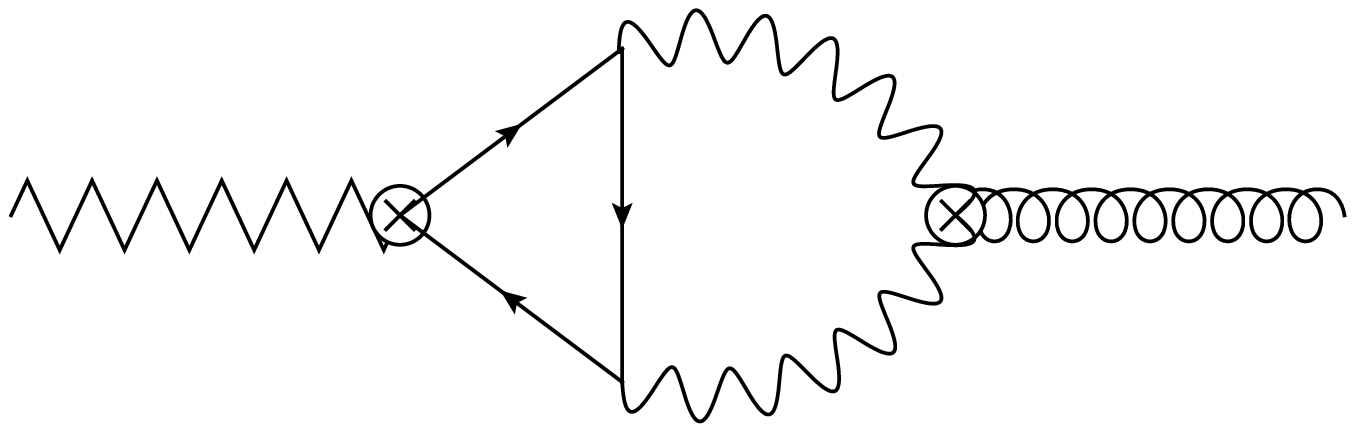}
			\caption{Leading order diagram.}
			\label{fig:2loop}
	\end{subfigure}
	\begin{subfigure}[b]{0.32\textwidth}
			\centering
			\includegraphics[width=\textwidth]{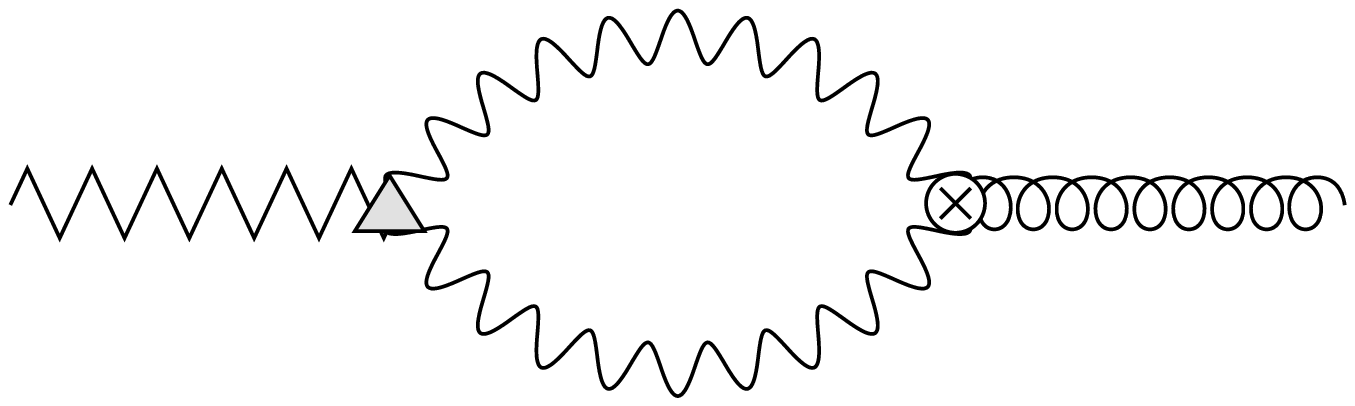}
			\caption{Effective leading diagram.}
			\label{fig:1loop}
	\end{subfigure}
  \caption{(a) The only class of diagrams with non-vanishing corrections to the CVE in the large $N$ limit. (b) The leading order diagram. (c) After replacing the leading order diagram with the effective vertex $J^\mu_\text{Anom}$. The triangle vertex represents the anomalous effect of the triangle diagram.}
  \label{fig:anom_diag}
\end{figure}

However, it turns out that there are certain diagrams in the second class that do in fact contribute to the CVE. Since in the absence of the anomaly in equation \eqref{eq:vector_ward}, there would be no radiative corrections, it is clear that the only contributing diagram are those that include a triangle subdiagram as in figure \ref{fig:AnomClass}.  Figure \ref{fig:2loop} shows the leading order diagram in this class which we now proceed to calculate.

Since we already know that the only contribution comes from the anomaly in \eqref{eq:vector_ward}, we can replace the triangle part of the two loop diagram with the effective vertex:
\begin{equation}
J^\mu_\text{Anom}=-\frac{g^2C(r)}{4\pi^2\sqrt{-g}}\e^{\mu\nu\rho\sigma} V_{\nu}\d_\rho V_{\sigma}
\label{eq:effective_vertex}
\end{equation}
which captures the divergence of the axial current, and we are effectively left with only a one loop diagram  (figure \ref{fig:1loop}). The calculation of the diagram is slightly complicated by the fact that different components of the gauge field receive different effective masses which are in general very important for our arguments. However, in this case it turns out that there are no infrared divergences even at zero effective mass and we present the calculation with this simplification.

The anomalous contribution to the Euclidean Green's function is:
\begin{equation}
\mathcal G^{i,0j}_\text{Anom}=\langle J^i_\text{Anom}T^{0j}\rangle=\frac{-g^2 T C(r)d(G)}{4\pi^2} \e^{ijm} k_m 
		\sum\hspace{-15pt}  \int \frac{d^3 p}{(2\pi)^3} 	\frac{\omega^2-\frac{p^2}3}{(\omega^2+p^2)^2}+\cdots
\label{eq:Anomalous_Greens}
\end{equation}
where $d(G)$ is the dimension of the adjoint representation and we have expanded up to first order in $k$. It is now a simple matter to carry out the sum and the integral:
\begin{equation}
\mathcal G^{i,0j}_\text{Anom}= \frac{ -g^2C(r)d(G)}{48\pi^4} \e^{ijm} k_m \int dp \, p\frac{1-e^{-2\frac{p}{T}}-4e^{\frac{-p}{T}}\frac{p}{T}}{\left(1-e^{\frac{-p}{T}}\right)^2}\to\frac{g^2C(r)d(G)}{48\pi^4} \e^{ijm} k_m,
\end{equation}
where in the last step we have removed the divergent zero temperature contribution before doing the integral. In fact, at zero temperature, this contribution is not allowed by full Lorentz symmetry and vanishes (This is also verified by direct calculation). Therefore, any regulator that respects full Lorentz symmetry in the zero temperature limit would reproduce the same result.

We now read off the leading order correction to the CVE coefficient in the large N limit:
\begin{equation}
\sigma^\mathcal V_{\small Correction}=\frac{g^2C(r)d(G)}{48\pi^2} T^2.
\end{equation}


\section{Connection To 3 Dimensional Chern-Simons Term}
\label{sec:CS}

The diagramatic arguments put forward in section \ref{sec:Coleman_Hill} were originally presented as proof of non-renormalization of the $U(1)$ Chern-Simons term in 3 dimensions. It is well known that the topological nature of the non-abelian Chern-Simons term leads to its quantization and therefore, does not receive radiative corrections.  Similar arguments for the Abelian case can be made if one assumes a compact spatial manifold. 

In this section, we will show that under similar assumptions, one can show that the quantization of the Abelian Chern-Simons term in 3-dimensions leads to the non-renormalization of the CVE coefficient. First, we show that gauge invariance leads to the quantization of the level and then, as an example, we verify that the free theory satisfies these quantization conditions.

\subsection{Quantization Of The Level}
In order to see the connection between the 3-dimensional Chern-Simons term and the CVE, we dimensionally reduce the theory on the compactified time direction\cite{Banerjee:2012}. To this end, we rename the relevant sources:
\begin{align}
a_i & = A_i(x),\\
b_i & = T g_{0i},
\end{align}
which transform under gauge and coordinate transformations with respective parameters $\alpha$ and $\e_\mu$ as follows:
\begin{align}
a_i & \to  a_i + \d_i \alpha,\\
b_i & \to  b_i + T \left(\nabla_i \e_0 + \nabla_0 \e_i\right),
\label{eq:gauge_trans}
\end{align}
where $\e_\mu$ is the diffeomorphism parameter ($x^\mu\to x^\mu+\e^\mu$). As we argued in section \ref{sec:Coleman_Hill}, these 3-dimensional gauge transformations are not anomalous, even in the presence of anomalies in the underlying 4-dimensional theory\footnote{For coupling to gauge fields, this is only true at leading order in $1/N$.}. The term of concern in the effective action is:
\begin{equation}
S_{eff}=i\,\k \int {d^3 x \e_{ijk}}a_i\d_j b_k=i\,\k T \int {d^4 x \e_{ijk}}a_i\d_j b_k,
\label{eq:Leff}
\end{equation}
where in the final equality we have inserted a spurious factor of $T$ along with a trivial time integral in order to make contact with the 4-dimensional 2 point function at hand. With this definition, the Green's function becomes:
\begin{equation}
\mathcal G^{i,0j}= T^2 \frac{\delta}{\delta a_i}\frac{\delta}{\delta b_j}Z_{eff}= -\k T^2 \e_{ijk} p_k,
\label{eq:CVE-quant}
\end{equation}
which leads to a contribution ot the CVE coefficient of 
\begin{equation}
\sigma^\mathcal V_A=\k T^2.
\label{eq:CVE_quant}
\end{equation}

To see the quantization of the coefficient $\k$, we put the system on a 3 torus of sides $L_1$, $L_2$ and $L_3$.  The periodicity of the manifold dictates the periodicty of the compact gauge fields $a_i$ and $b_i$. Using large gauge transformations $\alpha=\dfrac{2\pi}e \dfrac{x_i}{L_i}$ and $\e_0= \dfrac{2}{T} \dfrac{x_i}{L_i}$, we see:
\begin{align}
a_i & \equiv  a_i + \frac{2\pi}{e L_i},\\
b_i & \equiv  b_i + \frac{2}{L_i},
\label{eq:periodicity}
\end{align}
where $e$ is the electric charge and the factor 2 in the time translation parameter $\e_0$ is compensating for the anti-periodicity of fermions in the time direction.

We then perform a large gauge transformation on the $a_i$ fields with $\alpha = 2\pi n \dfrac{x_3}{e L_3}$, with $n \in \mathds Z$. The change in the action is:
\begin{equation}
\delta S_{eff}=2\pi n i \frac{\k}{e}  \int\limits_{x_3=0}\!\e_{3jk} \d_j b_k.
\end{equation}
Choosing a field configuration with non-trivial winding around the compact directions such as $b_2=\dfrac{2 x_1}{L_1 L_2}$, the variation becomes:
\begin{equation}
\delta S_{eff}=2\pi n i \frac{\k}{e}  \int\limits_{x_3=0}\!\e_{312} \d_1 b_2= 4\pi n i \frac{\k}{e},
\label{eq:Leff_var}
\end{equation}
which gives us the desired quantization result of $\k\in \dfrac{e}{2}\mathds Z$.\footnote{This is half the expected result of integer $\k$ values and is solely due to the fact that the fermions are anti-periodic in the thermal direction as stated under equation \eqref{eq:periodicity}.}

\subsection{Zero Coupling Example}
\label{sec:Free_coupling}
We now look at the Kaluza-Klein reduction of the action \eqref{eq:Fermion_L} at zero coupling. The fermions decompose on the thermal circle with periodicity $\beta=\frac1T$ as:
\begin{align}
\psi(t,x)=e^{2\pi i n \frac t T} \psi_{n}(x), \;\; n\in\mathds{Z}+\frac12,
\end{align}
and we assume the background sources have no time dependence. At zero coupling the different Matsubara frequencies do not mix and we can analyze the partition function mode by mode. For the $n$'th mode we have:
\begin{equation}
S=\frac{1}{T}\int d^3x \, \big(i\bar{\psi}_{-n}\cancel \d \psi_n + A_\mu J_{5,n}^\mu - h_{0i}T_n^{0i}\big),
\label{eq:Fermion_L_KK}
\end{equation}
with
\begin{align}
J_{5,n}^\mu=&\bar\psi_{-n}\g^\mu\g^5\psi_n,\notag\\
T^{0i}_n=&\frac{i}{4}\bar\psi_{-n}\g^0(\stackrel{\rightarrow}{\d^i}-\stackrel{\leftarrow}{\d^i})\psi_n
			-n\pi\bar\psi_{-n}\g^i\psi_n.
\end{align}
The diagram to be evaluated is given in figure \ref{fig:generic_diagram} with no internal vertices. We have:
\begin{align}
\mathcal G^{i,0j}=-T\int \frac{d^3 p}{(2\pi)^3}\text{Tr} \frac{\cancel{(p+k)}_3 -2\pi m T \g^0}{(p+k)^2+4\pi^2 m^2 T^2}\left(\frac{\g^0}{4T}(2p+k)^j-m\pi \g^j \right)
			\frac{\cancel p_3-2\pi m T \g^0}{p^2+4\pi^2 m^2 T^2}\g^i \g^5,
\end{align}
where $\cancel p_3$ denotes contraction with $\g^i$, the spatial gamma matrices. We have:
\begin{align}
\e_{ijn} \mathcal G^{i,0j}=&-T\int \frac{d^3 p}{(2\pi)^3}\frac{\e_{ijn}\text{Tr} (\g^0\g^r\g^s\g^i\g^5) N_{rs}^j}{((p+k)^2+4\pi^2 m^2 T^2)(p^2+4\pi^2 m^2 T^2)},
\end{align}
where
\begin{align}
N_{rs}^j=&(-2\pi mT)(-\pi m \delta^j_r)p_s-(p+k)_r \frac1{4T}(2p+k)^j p_s+(p+k)_r(-\pi m \delta^j_s)(-2\pi m T)\notag\\
=&-2\pi^2m^2 T \delta^j_r k_s-\frac1{4T}k_rp_s(2p+k)^j+ \text{ Sym in } r\leftrightarrow s.
\end{align}
Using the above and taking the limit of $k\to 0$, we have:
\begin{align}
\lim_{k\to0}\e_{ijn} \mathcal G^{i,0j}=& T\int \frac{d^3 p}{(2\pi)^3}\frac{\e_{ijn} 4 \e^{rsi}  (2\pi^2m^2 T \delta^j_r k_s+\frac1{2T}k_rp_s p^j)}
			{(p^2+4\pi^2 m^2 T^2)^2}.
\end{align}
This integral is of course divergent and needs to be regularized. Here we simply use the relation between different integrals in dimensional regularization\footnote{Here we just need the simplest of these relations: \[
\int \frac{d^dl}{(2\pi)^d}\frac{l^\mu l^\nu}{(l^2+\Delta)^2}=\frac{-\Delta}{d-2}\int \frac{d^dl}{(2\pi)^d}\frac{1}{(l^2+\Delta)^2}.\]}. Pauli-Villars regularization gives the same result:
\begin{align}
\lim_{k\to0}\e_{ijn} \mathcal G^{i,0j}=& 16 m^2 T^2 k_n \int\limits_0^\infty d p  \frac{p^2}{(p^2+4\pi^2 m^2 T^2)^2},\notag\\
						=&2\big|m\big|Tk_n.
\end{align}
And finally we write down the contribution to the CVE coefficient for each mode:
\begin{equation}
 \sigma^\mathcal V_{A,m}= \big|m\big| T.
\label{eq:}
\end{equation}
Which satisfies the quantization condition in the previous section. We evaluate the full CVE coefficient:
\begin{equation}
\sigma_{A}^\mathcal V = T \sum\limits_{m=-\infty}^\infty \sigma_{A,m}^\mathcal V=2 T^2\sum\limits_{m=\frac12}^\infty m \to \frac1{12} T^2.
\label{eq:CVE_full}
\end{equation}
Where we have used the $\zeta$-function regularization. This is the same result as \cite{Landsteiner:pert}. In essence, this calculation is the same with the order of the 3-dimensional momentum integral and the Matsubara sum.

\section{Conclusion}
\label{sec:concl}

We have shown in detail that, in so far as the perturbative loop expansion is valid, the zero chemical potential, temperature dependent part of the chiral vortical effect coefficient does not receive any radiative corrections from Yukawa type couplings.  We have also shown that this non-renormalization is related to the quantized nature of the 3-dimensional effective action.

None of this holds, however, when there are couplings to dynamical gauge fields present in the theory. In this case, even at large $N$,  radiative corrections are still present and we have calculated the leading correction to the CVE coefficient in this limit.

It is curious that the actual value of the CVE coefficient does not satisfy the quantization condition, despite the fact that the contribution from each Matsubara mode does. Considering the fact that the proof of the quantization requires invariance under \emph{both} large axial gauge and large diffeomorphism transformations, it seems very probable that this value would be fixed by considerations of global anomalies\cite{Witten:glob_curr,Witten:glob_grav}. This would also explain why arguments based on anomalous Ward identities such as in \cite{Jensen:2012jy}, which are consequences of anomalies only in the infinitesimal transformations, would not be able to determine this coefficient.

\subsection*{Acknowledgments}
This work is supported, in part, by DOE grant DE-FG02-00ER41132. The authors thank David B. Kaplan and Larry Yaffe for discussion. We are also greatful to Hai-cang Ren for pointing out an error in the first version of this paper.

\bibliographystyle{utphys}
\bibliography{biblio}

\end{document}